# PRELIMINARY EXPERIMENTS WITH EVA - SERIOUS GAMES VIRTUAL FIRE DRILL SIMULATOR

José Fernando M. Silva[1], João Emílio Almeida[1†], António Pereira[1†], Rosaldo J. F. Rossetti[1†], António Leça Coelho[2]

[1]Department of Informatics Engineering
[†]LIACC – Laboratory of Artificial Intelligence and Computer Science
Faculty of Engineering, University of Porto
Rua Roberto Frias, S/N, 4200-465, Porto, Portugal
{ei06123, joao.emilio.almeida, rossetti, amcp}@fe.up.pt

[2]LNEC – National Laboratory of Civil Engineering
Av. Brasil, 101, 1700-066, Lisboa, Portugal
alcoelho@lnec.pt

**ABSTRACT**

Fire keeps claiming a large number of victims in building fires. Although there are ways to minimize such events, fire drills are used to train the building occupants for emergency situations. However, organizing and implement these exercises is a complex task, and sometimes not sucessfull. Furthermore, fire drills require the mobilization of some finantial resources and time, and affect the normal functioning of the site where they occur. To overcome the aforementioned issues, computer games have a set of features that might overcome this problem. They offer engagement to their players, keeping them focused, and providing training to real life situations. The game evaluate users, providing them some feedback, making possible for the players to improve their performance. The proposed methodology aims to study the viability of using a game that recreates a fire drill in a 3D environment using Serious Games. The information acquired through the player's performance is very valuable and will be later used to implement an artificial population. A sample of 20 subjects was selected to test the application. Preliminary results are promising, showing that the exercise had a positive impact on users. Moreover, the data acquired is of great important and will be later used to demonstrate the possibility of creating an artificial population based on human behaviour.

**INTRODUCTION**

Statistics related to building fires reveals a high loss of lifes as result of them. In addition to the goods destroyed during a fire, there is still a high number of victims, some dead, others with severe injuries. Events such as the recent fire in the Brazilian discotheque "Kiss", this past January, where more than 230 young lives were lost, happen more frequently than they should. Some authors allege that failures during the evacuation process are one of the causes that mainly contributes for the building fires victims. Although there are ways to reduce the impact of this tragic events, fire keeps claiming a considerable number of victims. Emergency plans are poorly designed, and even worse implemented. In Madrid, during the Halloween party, an overcrowded concert led to the death of five young women when spectators crushed in one of the tunnels (illustrated in Figure 1).

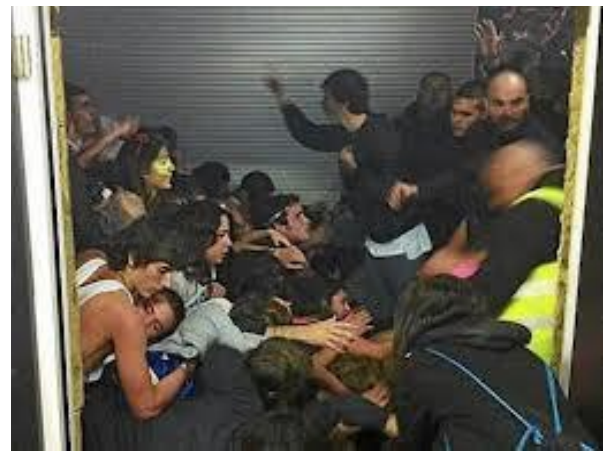

**Figure 1 - Stampede in Madrid Arena, 2012[1]**

Fire drills are used to train the building occupants for emergency situations. However, the participants some times have prior knowledge of their realization and, as result of this, they are not focused as they should. Moreover, fire drills require the mobilization of some resources, with the correspondent finantial costs: facilities must stop their normal activities, businesses are affected, educational activities in schools are compromised, surgeries and medical appointments must be rescheduled. Performing a fire drill will always affect the normal functioning of the place where they take place; for this reason, sometimes they are not performed at all. In some special locations, like hospitals, fire drills are unsuitable.

To overcome the aforementioned issues, computer games have a set of features that enable them to address this problem. In fact they provide engagement to their players, keeping them concentrated; besides they allow players to become experts in the resolution of challenges. The game allows evaluating the users' performance, providing also to them some feedback, and making possible for the players to improve their skills.

---

[1] http://www.20minutos.es/noticia/1641576/0/cronologia/tragedia/madrid-arena/

Some work has been done in this domain, proposing the use of Serious Games (SG) as a means to overcome such drawbacks, since immersion into the emergency scenario artificially created using computer videogames is easier to accomplish. The availability of game engines such as Unity3D provide a rapid way for prototyping 3D scenarios thus enabling to recreate the environment needed for a virtual fire drill simulator.

Another important issue is concerned with the acquisition of human behaviour in such emergency situations. By using the SG concept, it is possible to record some metrics regarding the players' decisions.

In this paper a test bed is presented, including the results obtained from a sample of 20 players. The metrics, and some data that were recorded, was later used to drive the artificial agents trying to recreate the players' decisions, based on their previous selections and the selected category of behaviour.

The remaining part of this paper is organised as follows. We start by briefly presenting some related concepts that concern this project, such as pedestrian evacuation simulators and serious games. We then discuss on applying serious games to evacuation training, following the presentation and formalisation of our problem. We propose the approach implemented in this paper and suggest a preliminary experiment using our prototype. Some results are also discussed, after which we finally draw some conclusions and give clues of some further steps in this research.

## BACKGROUND AND RELATED WORK

### Pedestrian evacuation simulators

Pedestrian evacuation simulators, a subset of pedestrian computer simulations, are developed mainly to test scientific theories and hypotheses, to assess design strategies, and to recreate the phenomena about which to theorize (Pan et al., 2007). Applications range from the entertainment to more serious uses like pedestrian behaviour in the real world or in panic situations (Almeida et al, 2011). Another important domain of application consists in the evaluation of the level of life safety provided by buildings (Kuligowski et al., 2010). When engineers use performance-based design approach to assess buildings' life safety, sometimes they use hand calculations, based on the equations provided by the Society of Fire Protection Engineers Handbook (SFPE, 2002), to provide them approximate egress times. However, these calculations are far from being accurate and other methodologies should be used instead (Coelho, 1997).

To correctly represent pedestrian flow, both the collective and the individual issues should be addressed (Hoogendoorn et al., 2004). Timmermans et al. (2009) states that the pedestrian decision-making process, as well as its movement, is of critical importance in the development of pedestrian models that aim to reproduce the reality. Teknomo divided pedestrian studies in two phases, namely data collection and data analysis (Teknomo, 2002). Whilst data collection focuses on characteristics such as speed, movement and path-planning, the latter is more focused in the pedestrians' behaviour.

Models can be classified according the level of depth: macroscopic or microscopic. Predicting the movement of crowds fall in the first category whilst individual pedestrian actions are in the second. The latter (microscopic level) is where most of pedestrian simulators are focused. Models of crowd based on the hydraulic or gas metaphor are used for the macroscopic level (Santos and Aguirre, 2004). In the microscopic models each person is described individually, thus allowing individual behaviours to be taken into account. According to (Castle, 2007), one possible classification of pedestrian evacuation simulators uses the occupants or enclosure perspective: i) coarse network models provide a macroscopic approach; ii) cellular automata models (Neumann, 1966, Beyer et al., 1985) and iii) continuous space models present a microscopic view (see Figure 2).

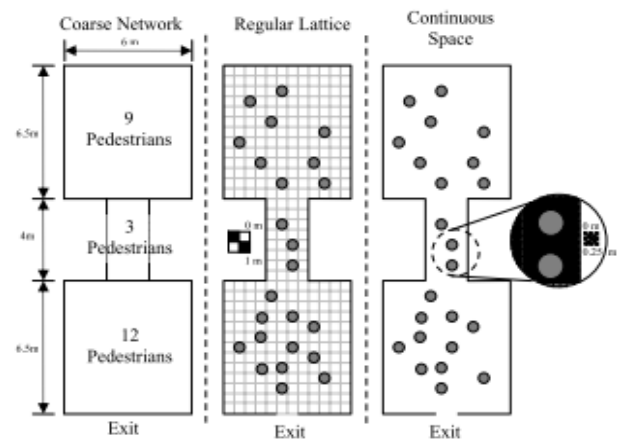

**Figure 2 - Examples of pedestrian models (Castle, 2007)**

Two of the various possible ways to model the pedestrian movement and interactions between persons, obstacles and the environment, are the magnetic forces model (Okazaki & Matsushita, 1993) and the social forces model (Helbing et al., 1995; Helbing et al., 2001; Helbing et al., 2002).

The validation and verification of pedestrian models relies on data collected by means of direct observations, photographs, time-lapse films (Coelho, 1997; Qingge et al., 2007) and also by stated preferences questionnaires (Cordeiro et al, 2011).

The emergent behaviour of groups of animals was addressed by Reynolds, who proposed a model to simulate the aggregate motion in a realistic way (Reynolds, 1987). Kuligowski had indeed studied human behaviour under emergency situations and she had also brought some insights on this matter regarding human behavioural process during evacuation from buildings (Kuligowski, 2008, 2011).

**Serious Games**

Video games were the leitmotif for advent of Serious Games (SG) with multiple applications other than entertainment, using appealing software with high-definition graphics and state-of-the-art gaming technology, for purposes that go beyond the traditional ones, such as educational as well as training. The variety of applications includes a wide range of domains, being social simulation one of them (Frey et al., 2007; McGonigal, 2011; Ribeiro et al., 2012).

SG purpose varies accordingly with the domain of application, such as education, training, health, advertising or social change (Hays, 2005).

Benefits of combining SG with other training activities include: the learners' motivation is higher; completion rates are higher; possibility of accepting new learners; possibility of creating collaborative activities; learn through doing and acquiring experience (Freitas, 2006).

The aforementioned aspects are a subset of the characteristics of SG-based frameworks that makes them so appealing to use for social simulations.

**EVA: A SG EVACUATION SIMULATOR**

Based on the Unity3D game engine, some research has been carried out at LIACC concerning the development of a SG-based Evacuation Simulator (Ribeiro et al, 2012). The version presented in this paper was coined EVA from the three initial letters of EVAcuation. It also relates to the first woman known, Eve (that is Eva in Latin languages, such as in Portuguese). Unity3D is a successful platform used worldwide for the development of video games, with appealing graphics, in which we can use the First Person Shooters (FPSs) game genre.

Unity3D was selected due to its characteristics among others: i) availability and free use; ii) powerful graphical interface; iii) ability to import models from other sources, such as Revit from Autodesk; iv) capability to develop code in JavaScript, C# or Boo. Detailed characteristics of the implemented environment are presented hereafter.

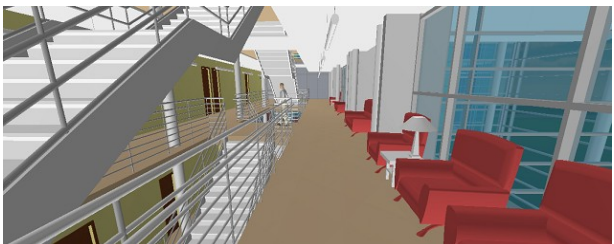

**Figure 3 - EVA screenshot**

**The game genre – First Person Shooter**

First Person Shooters (FPS) are used widely in video games. FPS are characterised by placing players in a 3D virtual world which is seen through the eyes of an avatar. Aim is to give the feeling of immersion in the virtual environment (illustrated in Figure 3).

The controls for this game use the FPS common standards, combination of keyboard and mouse to move the player around the environment. The complete action mapping is as follows:

- **Mouse movement** - camera control, i.e. where the player is looking at;
- **W** - move forward;
- **S** - move backwards;
- **A** - move to the left;
- **D** - move to the right;
- **F** – to open or close doors.

**Game scenario**

The game scenario used, a large three-floor building, was imported from Autodesk's Revit and its external view is represented in Figure 4.

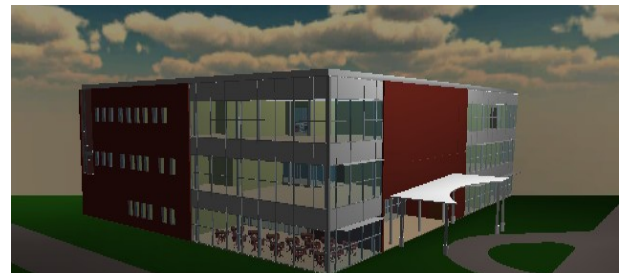

**Figure 4 - Exterior view of the building used**

The 3D model was converted into the Unity3D framework, and some furniture was added to increase the realms and give a more realistic, almost photo-quality, ambience. A character using the FPS-game genre was created. Mr. Adam is another character in a wheelchair, a patient in a clinic hospital (shown in Figure 5).

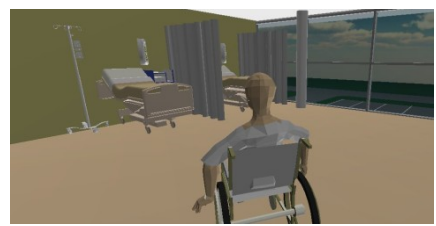

**Figure 5 - Inside view of Mr. Adam's Ward**

**Game description**

The game starts with the player at location "E" (see Figure 6), leaving the lift and pushing Mr. Adam's wheelchair towards his ward at "F". Suddenly, the fire alarm bells start ringing. The player is asked then what should be done (Figure 7). Possible answers include: a) nothing; probably it is a false alarm; b) wait for security personnel instructions; c) try to understand what is going on; d) leave the building as quickly as possible. The option selected is saved for later analysis.

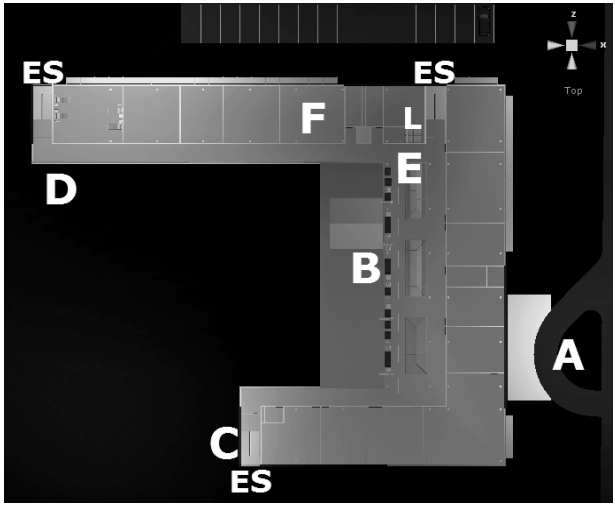

**Figure 6 - Building layout: A-main entrance; B-back entrance; C-south wing exit; D-northwest exit; E-starting point; F-ward; L-lift; ES-emergency exit**

Then, the player is instructed to leave the building due to a real fire alarm. Meanwhile, another important decision has to be made: will the player bring Mr. Adam or not? This is another option that will be recorded. If the player chooses to steer Mr. Adam in his wheelchair towards the exit reaching a safe zone, such as a different fire compartment or protected emergency stairs, the total time since the beginning of the fire alarm until that point is registered, and the player is greeted for that achievement, that is, for having rescued Mr. Adam to a safe zone.

Finally, the player is urged to go as quickly as possible to the outside to find a route to exit the building. The game will end as soon as the player reaches a valid exit. Total evacuation time is recorded. The valid exits are shown in Figure 4: A is building main entrance; B is back entrance or access to back yard; C is the exit of south wing emergency stair; D is the exit of northwest emergency stair.

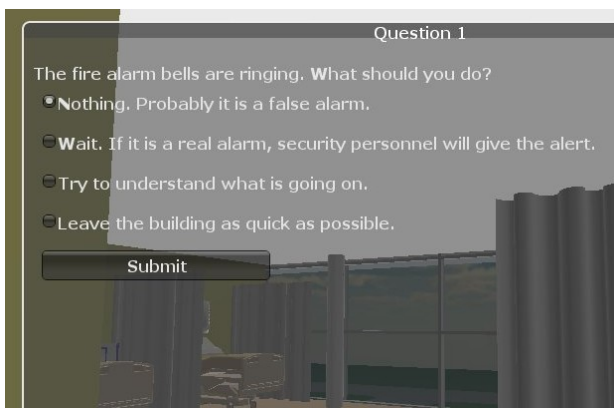

**Figure 7 - Questionnaire presented to the player after the fire alarm sounds**

### EXPERIMENTAL SETUP

The EVA prototype was run under a Windows 7 typical commodity computer with no special specifications. All subjects were tutored in the game controls usage, keyboard and mouse, and were given a small time to adapt on controlling the FPS character. Data acquired during the game was recorded in a text file. The players had no previous knowledge of the building or scenario and, to keep them the same chances, each player had only one run, to capture first reactions to the game experience and its controls, to avoid biased the data. This aspect is important to be noticed because one goal was to acquire genuine surprise with the fire alarm situation and to record the behaviour. For that reason, only one run was allowed for each player, because the next time he would be already expecting the alarm and what was expected to do.

### Population Sample

A total of 20 subjects were selected as sample to test the developed prototype. These testers can be classified according to the parameters presented in Table 1. This data was collected in a questionnaire presented to the players at the end of the game.

**Table 1 - Population sample**

|  | YES | NO |
|---|---|---|
| Regular video game player | 13 | 7 |
| Previous training in fire safety | 9 | 11 |
| Previous fire drill's experience | 12 | 8 |
| Been into a real fire | 1 | 19 |
| Followed emergency signage to find exit route | 16 | 4 |

### Description

As stated before, each subject could play only once. Some time was given to the user to get acquainted with the keyboard and mouse controls. No explanation on the game purpose, other than having to steer Mr. Adam in his wheelchair to the ward, was given. After hearing the alarm siren, the player had to answer the questionnaire shown in Figure 7. Finally, the player was advised to go as quickly as possible to the outside of the building. The game ends as soon as the player passes one of the building's exit.

### Data Collected

The results of answers to the initial questionnaire (Figure 7) are presented in Table 2. From the possible answers, the option c) or d) should be the ones to be chosen. If the subject, in a real situation, is confronted with a fire alarm, if he/she does not decided to leave immediately the building, at least should go and seek for information regarding the origin of such alarm. From the data collected, most players choosed options c) or d), equally, but two players choose the other options, one each.

**Table 2 - Answers to initial question**

|  | Answers |
|---|---|
| a) nothing; probably it is a false alarm | 1 |
| b) wait for security personnel instructions | 1 |
| c) try to understand what is going on | 8 |
| d) leave the building as quickly as possible | 9 |

Ignoring the alarm is a bad option and waiting for instructions from security personnel or some responsible person, is a risky option, because in a fire or emergency situation, things can go wrong and emergency plans might fail. Also the exit chosen by each player was registered, and the results are displayed in Table 3. Half of players opted for the back entrance, while only one opted for the south wing exit.

Table 3 - Exits used by the players

|  | Number of players |
|---|---|
| A-main entrance | 4 |
| B-back entrance | 10 |
| C-south wing exit | 1 |
| D-northwest exit | 5 |

**Preliminary results analysis**

The immersion provided by the game scenario was one of the positive remarks that most players made. The setup of the experiment was also referred to be a realistic one, concerning the goal of evaluating the reaction of users when faced with an unexpected fire alarm. Many of the players said that in a real situation their behaviour would be the same, thus proving a valuable feed back. For fire scientists and researchers, the issue of the pre-evacuation time, corresponding to the amount of time between the fire alarm until a decision is made, is of great importance. Not much data is available regarding this issue. Furthermore, this phase of the evacuation process is of crucial significance, because it can determine the success of the evacuation process.

It was noticed that some players did not choose the best or fastest route towards the outside of the building. Although the safest route and preferable one, from the starting point, was the northwest exit (D), only 5 of the subjects (25%) choose this exit. To this matter, the fact that many of the players confessed not to have followed the emergency signage, pointing towards the nearest and safest exit, must empahsized.

The evacuation times, although recorded, are not significant as absolute values, since the movement speed of the game character is set to 1.5 m/s and some aspects related with the cinematic movement of pedestrians, such as variable speed, walking versus running, and the steering of the characters, could bias the results. On the other hand, for validation and verification, experiments in the real building should be performed to compare results, but this is outside the scope of this experiment.

**CONCLUSION AND FUTURE WORK**

Since the population sample is reduced, results must be analysed with caution. However, the validity of the tool for training and to provide some education on how a building occupant should behave when confronted with a fire alarm, is promising.

Another aspect to be noticed is the possibility of using EVA and SG for human behaviour data collection. This data is of great importance to feed pedestrian simulators of evacuation scenarios. Particularly when this data can be divided

The very next steps in this research include the improvement of the prototype for rapidly setting up different simulation environments from Revit models of buildings. We also intend to include other performance measures to study individual and social behaviour in circumstances other than hazardous scenarios. Ultimately, this framework is also expected to be used as an imperative decision support tool, providing necessary and additional insights into evacuation plans, building layouts, and other design criteria to enhance places where people usually gather and interact rather socially.

**ACKNOWLEDGMENT**

This project has been partially supported by FCT (Fundação para a Ciência e a Tecnologia), the Portuguese Agency for R&D, under grant SFRH/BD/72946/2010.

**AUTHOR BIOGRAPHIES**

**JOSÉ FERNANDO MOREIRA DA SILVA** is a undergraduated student, will conclude his MSc in Informatics and Computing Engineering in 2013, from Faculty of Engineering, University of Porto, Portugal. He specialised in Digital Games development and Artificial Intelligence, combining the concepts of multi-agent systems and serious games. He can be reached by e-mail at: ei06123@fe.up.pt.

**JOAO EMILIO ALMEIDA** holds a BSc in Informatics (1988), and MSc in Fire Safety Engineering (2008). He is currently reading for a PhD in Informatics Engineering at the Faculty of Engineering, University of Porto, Portugal, and a researcher at LIACC. He has co-authored many fire safety projects for complex buildings such as schools, hospitals and commercial centres. His areas of interest include Serious Games, Artificial Intelligence, and multi-agent systems; more specifically he is interested in validation methodologies for pedestrian and social simulation models. His e-mail is joao.emilio.almeida@fe.up.pt.

**ROSALDO ROSSETTI** is an Assistant Professor with the Department of Informatics Engineering at the University of Porto, Portugal. He is also a Research Fellow in the Laboratory of Artificial Intelligence and Computer Science (LIACC) at the same University. Dr. Rossetti is a member of the Board of Governors of IEEE Intelligent Transportation Systems Society (IEEE ITSS) and a co-chair of the Technical Activities sub-committee on Artificial Transportation Systems and Simulation of IEEE ITSS. His areas of interest include Artificial Intelligence and agent-based modelling and simulation for the analysis and engineering of complex systems and optimisation. His e-mail is rossetti@fe.up.pt and his Web page can be found at http://www.fe.up.pt.com/~rossetti/.

**ANTONIO PEREIRA** was born in Porto, Portugal, and has a PhD in Informatics Engineering from University of Porto. Since 2003 he is dedicated to research in the field of Agent-Based Simulation, Distributed Artificial Intelligence, Intelligent Systems, and Optimization of Complex Systems. His e-mail is amcp@fe.up.pt and his personal webpage can be found at http://www.fe.up.pt/~amcp.

**A. LEÇA COELHO** holds both the Electrotechnical and Civil Engineering degrees, as well as a Master's and PhD in Civil Engineering. He is currently a Principal Researcher with Habilitation at LNEC. His areas of interest include fire safety and risk analysis. He can be reached by e-mail at `alcoelho@lnec.pt`.